\documentclass[conference]{IEEEtran}
\usepackage{graphicx}
\usepackage{amsmath}
\usepackage{amssymb}

\usepackage{cite}

\ifCLASSINFOpdf
\else
\fi

\usepackage{subfigure}

\begin{document}
%
\title{Cooperative Multiplexing in a Half Duplex Relay Network:
Performance and Constraints}

\author{\IEEEauthorblockN{Yijia Fan and H. Vincent Poor }
\IEEEauthorblockA{Dept. of Electrical Engineering\\
Princeton University\\
Princeton, NJ 08544\\
\{yijiafan, poor\}@princeton.edu} \and \IEEEauthorblockN{John S.
Thompson}
\IEEEauthorblockA{Institute for Digital Communications\\
School of Engineering and Electronics \\
University of Edinburgh \\
Edinburgh, United Kingdom, EH9 3JL \\
john.thompson@ed.ac.uk} }


%


\maketitle

\begin{abstract}
Previous work on relay networks has concentrated primarily on the
diversity benefits of such techniques. This paper explores the
possibility of also obtaining multiplexing gain in a relay network,
while retaining diversity gain. Specifically, consider a network in
which a single source node is equipped with one antenna and a
destination is equipped with two antennas. It is shown that, in
certain scenarios, by adding a relay with two antennas and using a
successive relaying protocol, the diversity multiplexing tradeoff
performance of the network can be lower bounded by that of a $2
\times 2$ MIMO channel, when the \emph{decode-and-forward} protocol
is applied at the relay. A distributed D-BLAST architecture is
developed, in which parallel channel coding is applied to achieve
this tradeoff. A space-time coding strategy, which can bring a
maximal multiplexing gain of more than one, is also derived for this
scenario. As will be shown, while this space-time coding strategy
exploits maximal diversity for a small multiplexing gain, the
proposed successive relaying scheme offers a significant performance
advantage for higher data rate transmission. In addition to the
specific results shown here, these ideas open a new direction for
exploiting the benefits of wireless relay networks.
\end{abstract}


%
\IEEEpeerreviewmaketitle

\section{Introduction}

Generally speaking, a relay network can act as a virtual
multiple-input multiple-output (MIMO) system if the nodes are
allowed to cooperate \cite{4,7,melda}. It is well known that a MIMO
system has two advantages over single-input single-output systems,
namely multiplexing gain and diversity gain. The diversity gain can
improve the system outage performance (i.e, reliability), while the
multiplexing gain enhances the spectral efficiency for high SNR. The
tradeoff between diversity and multiplexing gain is a key
characteristic of MIMO systems \cite{DMT,fDMT,holl}, and hence for
relay networks (virtual MIMO systems). The optimal
diversity-multiplexing trade-off (DMT) for half-duplex relay
networks is yet to be discovered \cite{7,melda}, especially in the
scenario in which multiple antennas can be deployed at one node.
However, instead of looking at both multiplexing and diversity
behavior simultaneously, most of the past work emphasizes primarily
the diversity benefits of the relay network(e.g.\cite{4}), while
ignoring the possible multiplexing benefits it could bring. Unlike a
point-to-point MIMO link, in a half-duplex relay network,
multiplexing gain is difficult to obtain due to the additional
transmission time slots the relays require. In fact, it has been
shown recently \cite{melda} that no multiplexing gain (of more than
1) can be achieved \emph{for high SNR in general}, when the source
is deployed with only one antenna, even if \emph{full-duplex} relay
transmission is assumed. We note that, compared with full-duplex
relaying, half-duplex relaying is recognized to be a suboptimal but
more practical choice for wireless networks.

As one might hope that relaying could bring both diversity and
multiplexing gain, investigating and realizing this possibility is
of significant importance. Very recently, some capacity analyses
\cite{itw06,ah} on \emph{scalar} \emph{channels} have shown that
\emph{only under certain signal to noise ratio (SNR) constraints},
is it possible to achieve a MIMO rate through \emph{full-duplex}
relaying. However, the DMT for these SNR values in fading
environments is not exploited and discussed in these papers.

In this paper, we show that it is even possible to obtain
multiplexing gain in a \emph{half-duplex} relay network. We consider
a scenario in which the relays perform decode-and-forward.
Specifically, we consider a one-antenna source, a two-antenna relay,
and a two-antenna destination. We apply a \emph{successive relaying}
protocol to make the two antennas at the relay transmit in turn. We
show that in this scenario a DMT that is at least as good as that of
a $2 \times 2$ MIMO channel can be obtained under certain finite SNR
or channel constraint. Based on our network model, we show that the
constraint can be expressed by an upper bound for the SNR and a
function of the channel coefficients. We also show that the above
DMT can be achieved with a very high probability for most of the
realistic SNR values, in a scenario in which the relay is close to
the source. We also develop a more practical signalling method,
which we refer to as the distributed D-BLAST architecture, to
achieve the $2 \times 2$ MIMO DMT lower bound. Furthermore, we
derive a space-time coding scheme, which can also offer a
multiplexing gain of more than $1$, provided that the source to
relay channel is good enough. We discuss the constraints for this
scheme and compare it with the successive relaying scheme. While the
space-time coding strategy exploits maximal diversity for a small
multiplexing gain, the successive relaying scheme offers significant
performance advantages for higher data rate transmission (i.e,
higher multiplexing gain).

The decode-and-forward successive relaying scheme has been discussed
for single antenna relay networks \cite{se1,fan2}, while neither of
the above works explore the possibility of obtaining multiplexing
gains of more than 1 by using such a scheme. We note that the
difference between our work and \cite{se1,fan2} is that we use
\emph{independent Gaussian codebooks} at the relays to re-encode the
message, instead of using the same codebook as at the source. In
fact, in our work the additional multiplexing gain is obtained
through \emph{distributed coding}.

The rest of the paper is organized as follows. In Section II, the
system model and transmission protocol are introduced. In Section
III, the DMT performance for the proposed scheme is analyzed. In
Section IV, a distributed D-BLAST signalling method is proposed to
approach the DMT bound obtained in Section III. The space-time
coding scheme is discussed and compared with the proposed successive
relaying scheme in Section V, and conclusions are drawn in Section
VI. Due to limited space, we omit all the proofs of the theorems in
the paper. Details of the proofs can be found in \cite{fancom}.

\begin{figure}[t!]
\centering
\includegraphics[width=3in]{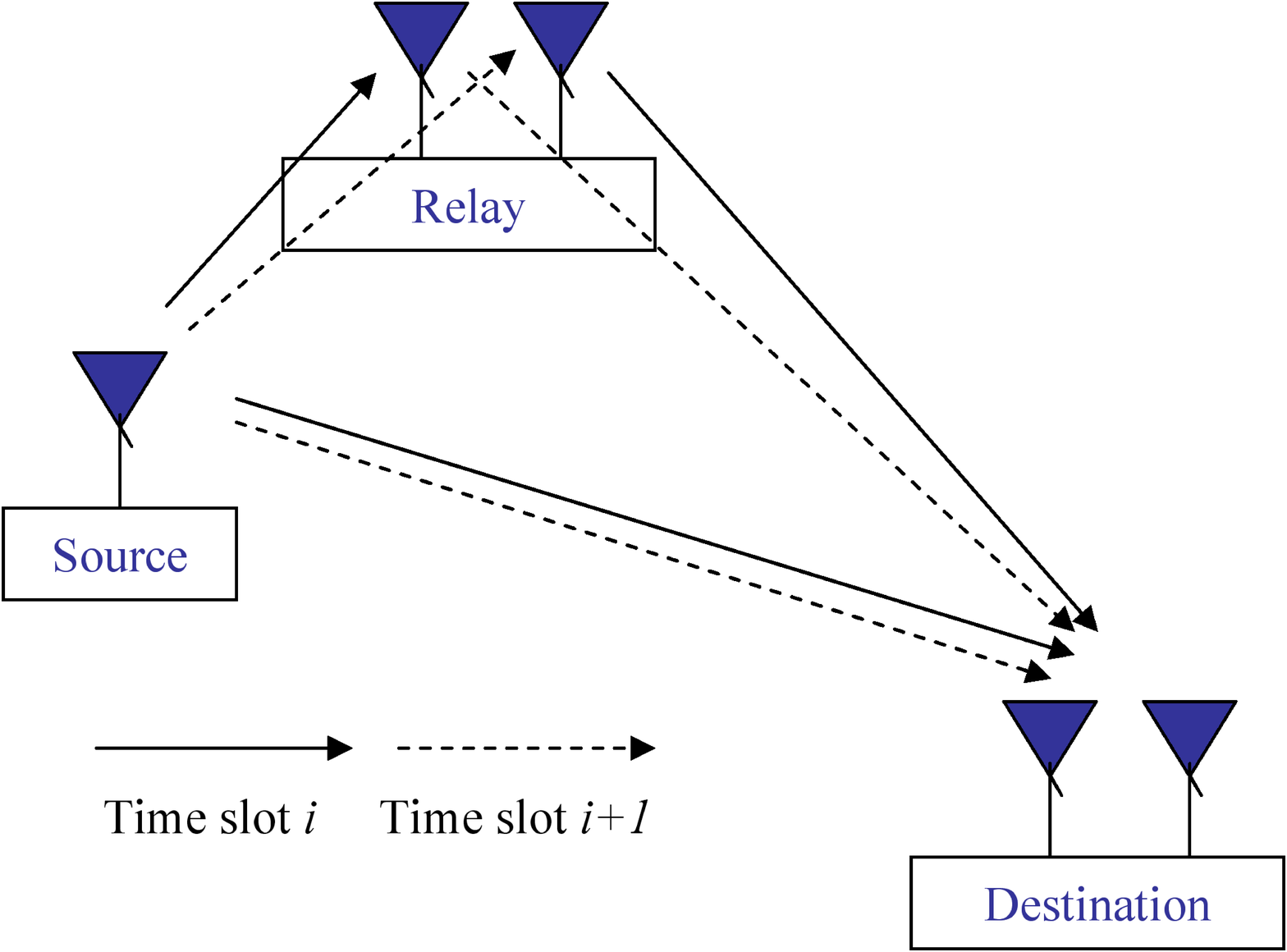}
\caption{Network model.} \label{parcod1}
\end{figure}

\section{System Model and Protocol Description}

We concentrate on a network in which there is one source having a
single antenna, one relay having 2 antennas, and one destination
having 2 antennas. We assume that the relay is close to the source,
while both the source and the relay are far away from the
destination\footnote{More specific examples will be given and
analyzed in section III and IV.}. Note that this assumption is made
to facilitate the decode-and-forward relaying protocol. We split the
source transmission into frames, each containing $L$ codewords
denoted as $x_l$ ($l=1,2,...,L$). Each $x_l$ represents a different
message. These $L$ codewords are transmitted continuously by the
source, and are decoded, re-encoded and forwarded by two antennas at
the relay successively in turn. When the relay re-encodes the
message, it uses a Gaussian codebook independent of the one used by
the source. For example, in time slot $i$, the source transmits a
codeword $x_i$, the destination and antenna $1$ at the relay receive
this codeword, while antenna $2$ transmits a codeword $x'_{i - 1}$,
which represents the same message as the codeword $x_{i-1}$ it
received in the previous time slot $i-1$ but which is generated from
an independent Gaussian codebook; in time slot $i+1$, the source
transmits another codeword $x_{i+1}$ to the destination and antenna
$2$ at the relay, while antenna $1$ at the relay transmits a
codeword $x'_i$, which represents the same message as $x_i$ but
generated from an independent Gaussian codebook, etc. Overall, $L$
codewords are transmitted in $L+1$ time slots. A visual description
for the network model and protocol is shown in Fig. 1. Note that in
each time slot $i$ ($2 \le i \le L$) at the relay, one of the
antennas is transmitting while the other is receiving. Here we make
an important assumption that the interference between the antennas
can be pre-subtracted, as the knowledge of the transmitted signal is
known to both antennas at the relay.

\section{DMT Analysis}

We assume a slow, flat, block fading environment, in which the
channel remains static for each message frame transmission (i.e.
$L+1$ time slots). We denote the source as $s$, relay as $r$,
destination as $d$, and the channel coefficient between the $i$th
transmit antenna at node $a$ and the $j$th antenna at the node $b$
by $h_{a_i,b_j }$. Each $h_{a_i,b_j}$ experiences independent and
identically distributed (i.i.d.) Rayleigh fading. We will also
consider other fading effects specifically later, such as path loss
in Section III.B. Unless specifically stated, we assume that the
transmitters do not know the instantaneous channel state information
(CSI) on their corresponding forward channels, while CSI is
available at the receivers on their receiving channels. We also
assume that all transmit antennas transmit with equal power
$\eta$\footnote{Note that this assumption is made for the sake of
analytical simplicity; any power allocation will affect the system
power gain but not the diversity multiplexing tradeoff at high
SNR.}. The white Gaussian noise at the receive antennas is assumed
to be i.i.d. with zero mean and unit variance. First, we review the
definition of the high-SNR DMT.
\newtheorem{definition}{Definition}
\begin{definition}
\textbf{(High-SNR DMT)} Consider a family of Gaussian codes $C_\eta$
operating at SNR $\eta$ and having rates $R$. Assuming a
sufficiently long codeword, the multiplexing gain and diversity
order are defined as
\begin{equation}
r \buildrel \Delta \over = \mathop {\lim }\limits_{\eta  \to \infty
} \frac{{R}}{{\log _2 \eta }},\ \ \mathrm{and} \ \ {\rm{ }}d
\buildrel \Delta \over = - \mathop {\lim }\limits_{\eta  \to \infty
} \frac{{\log _2 P_{out} \left( R \right)}}{{\log _2 \eta }},
\end{equation}
where ${P_{out}}\left( R \right)$ denotes the outage probability as
a function of the transmission rate $R$.
\end{definition}
We also review the maximal DMT that can be obtained \emph{in
general} for the system model described in Section 2.
\newtheorem{theorem}{Theorem}
\begin{theorem}[\cite{melda}]
The maximal DMT for the system model described in Section II (i.e,
one single-antenna source, one two antenna relay and one two antenna
destination) in half-duplex mode is $d\left( r \right) = 4\left( {1
- r} \right)^ +$.
\end{theorem}
We note that this bound might be achieved by using a
compress-and-forward protocol \cite{melda}, which is distinct from
the decode-and-forward protocol considered here.

\begin{figure*}[!t]
\begin{eqnarray}
\underbrace {\left( {\begin{array}{*{20}c}
   y_1^1   \\
   y_1^2   \\
   y_2^1   \\
 y_2^2  \\
 y_3^1  \\
 y_3^2  \\
  \vdots  \\
 y_{L+1}^1  \\
 y_{L+1}^2  \\
\end{array}} \right)}_{\bf{y}} = \sqrt \eta  \underbrace {\left( {\begin{array}{*{20}c}
   {h_{s,d_1 } } & 0 & 0 & 0 & 0 & 0 & 0  \\
   {h_{s,d_2 } } & 0 & 0 & 0 & 0 & 0 & 0  \\
   0 & {h_{r_1 ,d_1 } } & {h_{s,d_1 } } & 0 & 0 & 0 & 0  \\
   0 & {h_{r_1 ,d_2 } } & {h_{s,d_2 } } & 0 & 0 & 0 & 0  \\
   0 & 0 & 0 & {h_{r_2 ,d_1 } } & {h_{s,d_1 } } & 0 & 0  \\
   0 & 0 & 0 & {h_{r_2 ,d_2 } } & {h_{s,d_2 } } & 0 & 0  \\
   0 & 0 & 0 & 0 & 0 &  \ddots  & 0  \\
   0 & 0 & 0 & 0 & 0 & 0 & {h_{r_{1(2)} ,d_1 } }  \\
   0 & 0 & 0 & 0 & 0 & 0 & {h_{r_{1(2)} ,d_2 } }  \\
\end{array}} \right)}_{\bf{H}}\underbrace {\left( \begin{array}{l}
 x_1  \\
 x_1^{'}  \\
 x_2  \\
 x_2^{'}  \\
 x_3  \\
 x_3^{'}  \\
  \vdots  \\
 x_L  \\
 x_L^{'}  \\
 \end{array} \right)}_{\bf{x}} + \underbrace {\left( \begin{array}{l}
 n_1^1  \\
 n_1^2  \\
 n_2^1  \\
 n_2^2  \\
 n_3^1  \\
 n_3^2  \\
  \vdots  \\
 n_{L+1}^1  \\
 n_{L+1}^2  \\
 \end{array} \right)}_{\bf{n}},
 \label{io}
\end{eqnarray}
\hrulefill \vspace*{4pt}
\end{figure*}

\subsection{A Lower Bound for the Optimal DMT}

In the following we will first assume that the relay can always
correctly decode the message. We will remove this assumption later
when considering the constraints on the source-relay link in Section
III.B. After the relay correctly decodes the message, it forwards
the re-encoded symbol to the destination. The destination
\emph{waits} \emph{until} it receives the entire message frame (i.e,
$L$ symbols) in $L+1$ time slots, before it performs maximal
likelihood (ML) decoding. The system input-output relationship, for
each set of $L+1$ time slots, can be expressed as equation
(\ref{io}), where the vector ${\bf{y}}$ is the $2(L+1) \times 1$
receive vector, the matrix $\bf{H}$ denotes the $2(L+1) \times 2L$
channel transfer matrix, $\bf{x}$ is the $2L \times 1$ transmit
vector, and $\bf{n}$ is the $2(L+1) \times 1$ noise vector at the
receiver. The scalar $y_i^j$ denotes the signal received by antenna
$j$ in the $i$th time slot. The scalar $x_i$ denotes the symbol
transmitted by the source in the $i$th time slot, while $x_i^{'}$
denotes the symbol transmitted by the relay in the $(i+1)$st time
slot, which contains the same message as in $x_i$. The channel
coefficients $h_{r_{1(2)} ,d_i }$ in the bottom-right of $\bf{H}$
could be either $h_{r_{1} ,d_i }$ (for odd $L$) or $h_{r_{2} ,d_i }$
(for even $L$). Finally, the scalar $n_i^j$ denotes the received
Gaussian noise at the $j$th antenna at the destination in the $i$th
time slot. Regarding the DMT, we have the following theorem for this
system.
\begin{theorem}
For sufficiently large $L$, the DMT for the system described in
(\ref{io}) is lower bounded by that for a $2 \times 2$
point-to-point MIMO channel, i.e, the piecewise linear function (see
Theorem 2 in \cite{DMT}) connecting the points $\left( {n,\left( {2
- n} \right)^2 } \right)$, $n = 0,1,2.$
\end{theorem}
This result is surprising, as it implies that MIMO multiplexing gain
can be obtained with only one (source) transmit antenna. The way to
interpret this is that the multiplexing gain in this scenario is in
fact obtained through distributed coding of the same information
across both space and time. In this way, every piece of information
is \emph{multiplexed} into two separate (independently) encoded data
streams in each of two successive time slots, while those two
streams are received by two antennas (at the destination). This
implies that a multiplexing gain of $2$ can be obtained, once a
reliable link between the two transmit antennas is established. The
D-BLAST structure to be introduced later will give an operational
interpretation for this system.

\subsection{Finite SNR DMT and Constraints}

When we consider the source to relay channel constraint, we consider
the more recent concept of the finite-SNR DMT, which allows us to
study the DMT for realistic SNRs. The definition of the finite DMT
is given as follows \cite{fDMT}.
\begin{definition}
\textbf{(Finite-SNR DMT)} The finite-SNR multiplexing gain $r$ and
diversity gain $d$ are defined as
\begin{eqnarray}
r = \frac{R}{{\log _2 \left( {1 + g\eta } \right)}}, \ \
\mathrm{and} \ \ d\left( {r,\eta } \right) =  - \eta \frac{{\partial
\ln P_{out} \left( {r,\eta } \right)}}{{\partial \eta }}
\end{eqnarray}
where $g$ denotes an array gain achieved at low SNR, and $P_{out}
\left( {r,\eta } \right) = P$ is the outage probability at rate $R =
r\log _2 \left( {1 + g\eta } \right)$.
\end{definition}

\begin{theorem}
Suppose the SNR satisfies the following constraint.
\begin{equation}
\eta  \le \frac{{a - b - c}}{bc}, \label{cond}
\end{equation}
where
\begin{eqnarray}
a&=&\min \left\{ \left| {h_{s,r_i } } \right|^2 \right\} \nonumber \\
b&=&\left( {\left| {h_{s,d_1 } } \right|^2  + \left| {h_{s,d_2 } }
\right|^2 } \right) \nonumber \\
c&=&\min \left\{ {\left| {h_{r_i ,d_1 } } \right|^2 + \left| {h_{r_i
,d_2 } } \right|^2 } \right\} \nonumber
\end{eqnarray}
for $i=1$ and $2$. Then, the outage probability for the network
shown in Section II is lower bounded by
\begin{equation}
P_{out}  \le 2P_{2 \times 2}-P_{2 \times 2}^2, \label{lob}
\end{equation}
where $P _{2 \times 2}$ is the outage probability for a $2 \times 2$
MIMO channel. The finite-SNR DMT for the proposed scheme performs
approximately the same as that for a $2 \times 2$ point-to-point
MIMO channel.
\end{theorem}

There are two important notes about this corollary.

\subsubsection{Impact of network geometry} One might think that the
probability with which (\ref{cond}) holds is low in general. This is
true. However, as widely indicated in the previous literature, it is
well recognized that the decode-and-forward protocol performs well
only when the source and relay are close to each other (e.g.
\cite{ah}). Otherwise other relaying modes might be better choices
(e.g. amplify-and-forward and compress-and-forward). In practice,
the best way to perform decode-and-forward is to use a relay that is
close enough to the source so that reliable communication between
the source and relay can be established.

Now, if we take the distance into account and consider the path
loss\footnote{Note that adding path loss does not affect the DMT
performance in general.}, it can be seen that (\ref{cond}) can be
satisfied with high probability in many scenarios. Suppose, for
example, that the distance between the source and the relay is
${\tilde r}$, while both source and relay are at unit distance from
the destination (note that similar models have been used in previous
analyses, e.g. \cite{melda,itw06} ). With a pathloss exponent of 4,
which is commonly assumed in a terrestrial environment, the channel
coefficients for source-relay links can be rewritten as ${\tilde
r}^{-2}{h_{s,r_i}}$. The constraints (\ref{cond}) can be
approximated as
\begin{equation}
\eta  \le {\tilde r}^{-4}  \times \frac{a}{bc}. \label{condpa}
\end{equation}
It can be seen that as long as $\tilde r$ is sufficiently small, the
probability that (\ref{condpa}) is satisfied is high. Simulation
results relating to this probability are described in Section V.

\subsubsection{Adaptive protocol design} The result offers many
insights for practical adaptive protocol design. For example, if the
relay (near the source) can be configured to retransmit the message
only when it decodes the signal correctly, then a MIMO DMT can be
obtained with high probability. Note that although it has been shown
that a MIMO DMT is not possible for the network model shown in the
paper \emph{in general}, the proposed transmission and coding scheme
offers MIMO DMT performance in many scenarios. Those scenarios are
\emph{not uncommon}, especially in an ad hoc network or uplink
cellular environment.

\begin{figure}[t!]
\centering
\includegraphics[width=3in]{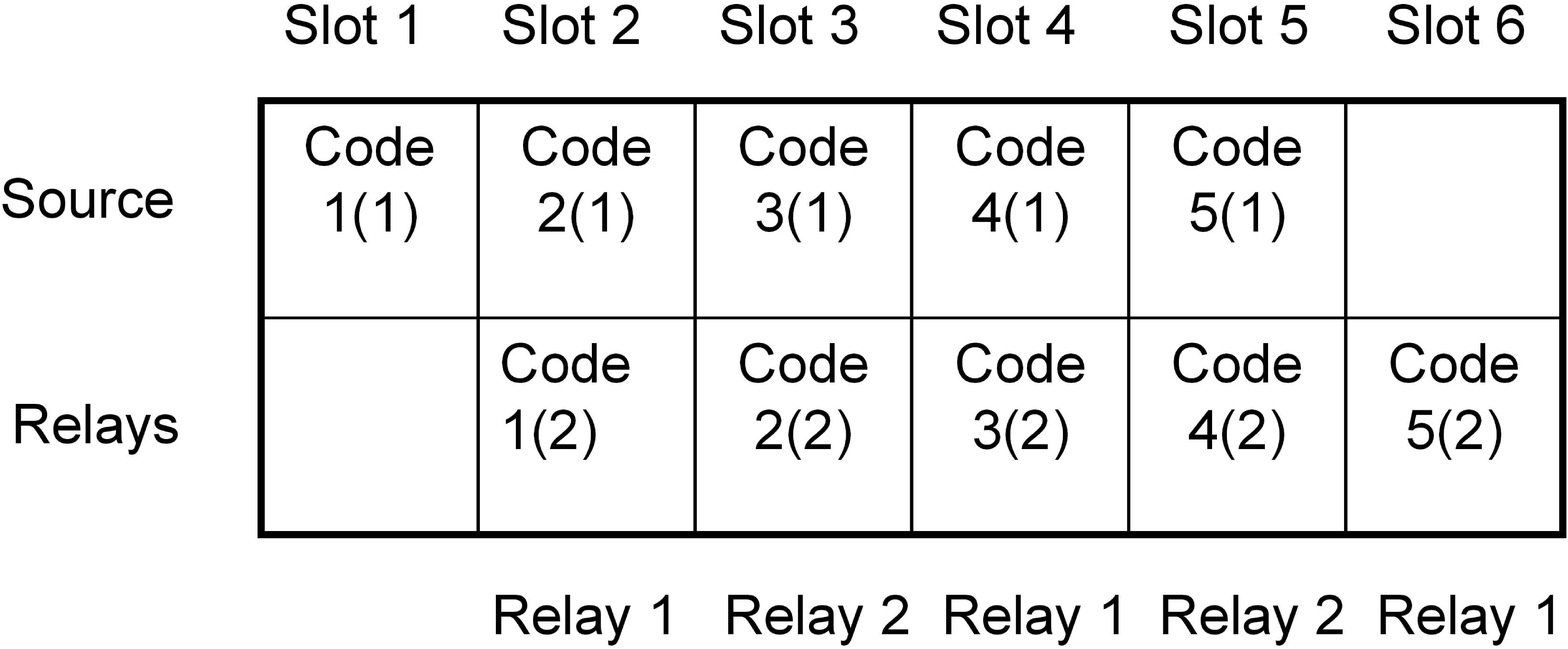}
\caption{Transmission schedule.} \label{parcod1}
\end{figure}

\section{Distributed D-BLAST}

The DMT-tradeoff presented in \emph{Theorem 2} might be obtained by
using an Maximum-likelihood (ML) decoder at the destination, which
is extremely complex for large values of $L$. A simpler decoder is
thus desired. Carefully examining the proposed transmission and
coding strategy, it can be found that the codes representing the
same messages are layered diagonally in space and time. Fig. 2
illustrates this structure, in which codeword $i\left( j \right)$
denotes the $j$th code for the $i$th message. Clearly this structure
mimics a D-BLAST structure, which was originally proposed for
point-to-point MIMO \cite{dblast}. The only difference is that the
messages are encoded in a distributed fashion at both the source and
the relay. It has been shown that D-BLAST with minimal mean-squared
error (MMSE) successive interference cancellation (SIC) is
DMT-optimal for a point-to-point MIMO system \cite{DMT}. We will
also show here that the distributed D-BLAST can achieve the same DMT
as indicated in \emph{Theorem 2}.

\begin{theorem}
For the system described in (\ref{io}), the distributed D-BLAST
structure with the MMSE SIC algorithm achieves the optimal DMT lower
bound as indicated in \emph{Theorem 2}.
\end{theorem}

Generally speaking, the performance advantage of the distributed
D-BLAST lies in the use of parallel channel coding methods (e.g. the
use of independent codebooks in the above analysis) for each
message. We note that recently developed parallel channel codes for
point-to-point MIMO systems might be applied directly to the
distributed D-BLAST structure here (e.g., see the bit-reversal
permutation codes in \cite{pcd}).

\section{Comparison with Space-time Coding}

This scheme can be considered to be an extension of Laneman's scheme
\cite{4} to a multiple antenna scenario, and we term it the
\emph{space-time coding protocol}. In this protocol, transmission is
divided into two time slots. In the first time slot, the source
broadcasts the signal to the relay and destination. Each antenna at
the relay uses an independent codebook to re-encode the message it
received, and transmits the new codeword to the destination in the
second time slot. Note that two independent codebooks are used at
the relay in total. Assuming that the relay correctly decodes the
message, the input-output relationship can be expressed as
\begin{equation}
\left( \begin{array}{l}
 y_1^1  \\
 y_1^2  \\
 y_2^1  \\
 y_2^2  \\
 \end{array} \right) = \eta \left( {\begin{array}{*{20}c}
   {h_{s,d_1} } & 0 & 0  \\
   {h_{s,d_2} } & 0 & 0  \\
   0 & {h_{r_1 ,d_1 } } & {h_{r_2 ,d} }  \\
   0 & {h_{r_1 ,d_2 } } & {h_{r_2 ,d} }  \\
\end{array}} \right)\left( {\begin{array}{*{20}c}
   x  \\
   {x^{'} }  \\
   {x^{''} }  \\
\end{array}} \right) + \left( \begin{array}{l}
 n_1^1  \\
 n_1^2  \\
 n_2^1  \\
 n_2^2  \\
 \end{array} \right)
 \label{stc},
\end{equation}
where $x^{'}$ and $x^{''}$ denote the two codewords at the relay. We
have the following theorem in terms of DMT for the system described
in (\ref{stc}).
\begin{theorem}
On assuming that the relay correctly decodes the message, the
high-SNR DMT for the system in (\ref{stc}) can be written as
\begin{eqnarray}
d\left( r \right) = \left\{ \begin{array}{l}
 6 - 6r,0 \le r \le 1/2 \\
 5 - 4r,1/2 \le r \le 1 \\
 3 - 2r,1 \le r \le 3/2 \\
 \end{array} \right..
\end{eqnarray}
\end{theorem}
This theorem shows that multiplexing gains greater than $1$ can be
obtained through space-time coding if multiple antennas are deployed
at the relay. We note that this fact was not discovered in \cite{4},
which only considers a single antenna network. In fact, it was shown
in \cite{4} that the network suffers from a multiplexing loss (i.e,
a multiplexing gain of less than 1) when the destination is deployed
with only a single antenna, even if the message is correctly decoded
at the relays. Therefore, we can conclude that deploying a single
antenna at the destination is not sufficient to fully exploit the
benefits of space-time coding. \emph{Theorem 5} also indicates that
the space-time coding scheme can outperform the DMT upper bound (in
\emph{Theorem 1}) that can be achieved \emph{in general}, \emph{as
long as} the message is correctly decoded at the relay.
Specifically, it can achieve a maximal diversity gain of $6$ for
zero multiplexing gain. However, in practice this assumption imposes
constraints on SNR as well as source-relay channel conditions.
\begin{figure}[t!]
\centering
\includegraphics[width=3.2in]{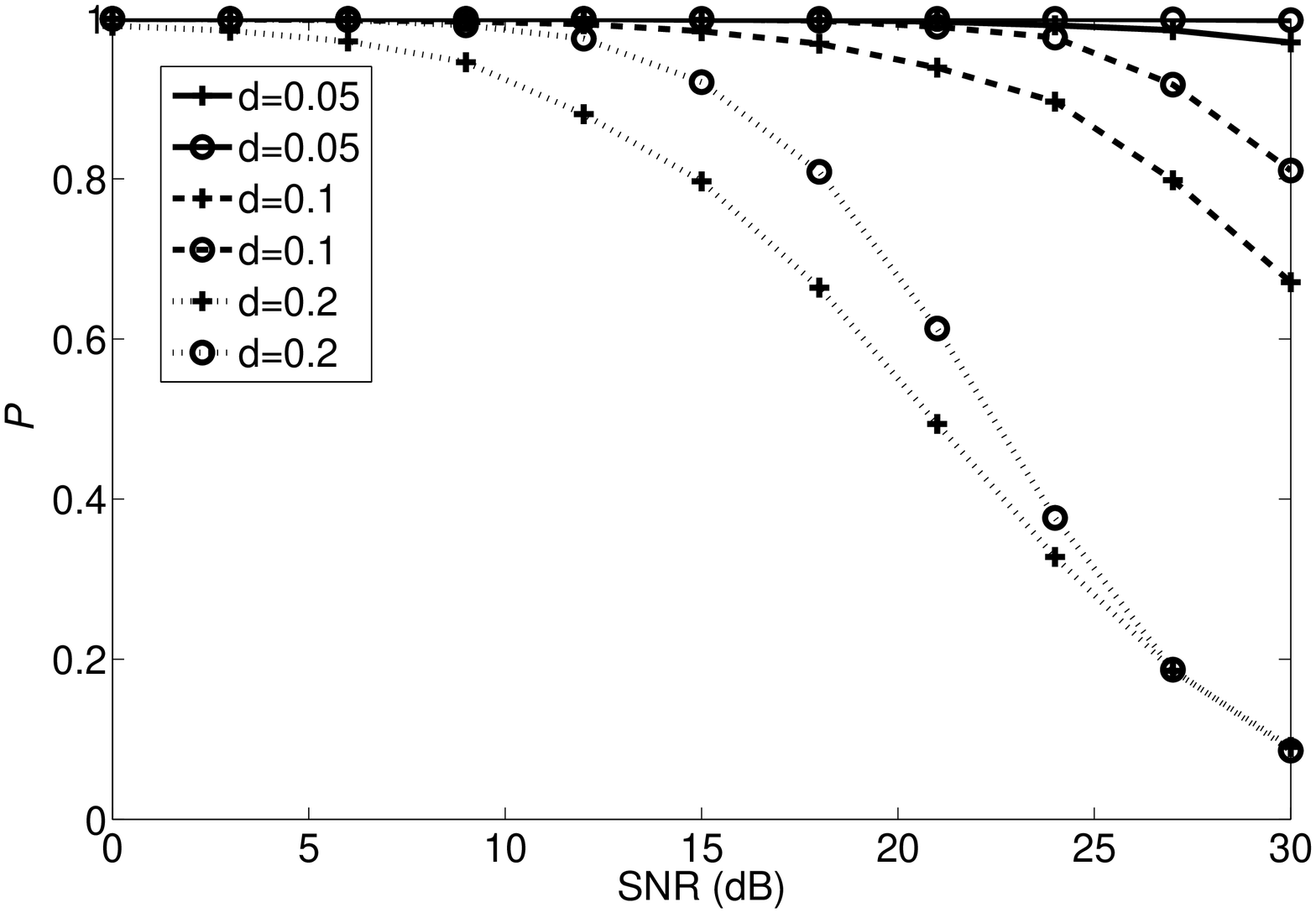}
\caption{The probability for (\ref{condpa}) and (\ref{cons2}) versus
SNR ($\eta$). The letter ``d'' denotes the distance between the
source and relay (i.e., $\tilde r$). The curves with mark ``+''
denote the performance of the successive relaying scheme. The curves
with mark ``o'' denote the performance of space-time coding.}
\label{dmfpc1}
\end{figure}
Using the same channel model as indicated in Section III.B, these
constraints can be approximated by the following bound (see
\cite{fancom} for details of the analysis):
\begin{equation}
\eta  \lesssim \tilde r^{ - 4}  \times \frac{p}{qz}, \label{cons2}
\end{equation}
where $p=\left| {h_{s,r_1 } } \right|^2 + \left| {h_{s,r_2 } }
\right|^2$, $q=\left( {\left| {h_{s,d_1 } } \right|^2  + \left|
{h_{s,d_2 } } \right|^2 } \right)$ and $z=\left( {\left| {h_{r_1
,d_1 } } \right|^2  + \left| {h_{r_1 ,d_2 } } \right|^2  + \left|
{h_{r_2 ,d_1 } } \right|^2 + \left| {h_{r_2 ,d_2 } } \right|^2 }
\right)$.

Fig. 3 shows the probability with which (\ref{condpa}) and
(\ref{cons2}) hold for different value of $\eta$. It can be seen
that when the source-relay distance is small (e.g. $\tilde r =0.05$,
or $\tilde r=0.1$ in the figure), the probability for both
constraints is high even for an SNR value of 30dB, which is higher
than in most practical applications. Note that the DMT performance
in this SNR region approaches the high-SNR DMT as indicated in
\emph{Theorem 2} and \emph{Theorem 5}. For a medium SNR level (e.g.
0-15dB), the probabilities approach 1. This means that the
finite-SNR DMTs for the systems in (\ref{io}) and (\ref{stc}) are
almost \emph{always} reached.

\begin{figure}
\subfigure[$r=1.$]{\includegraphics[width=3.2in]{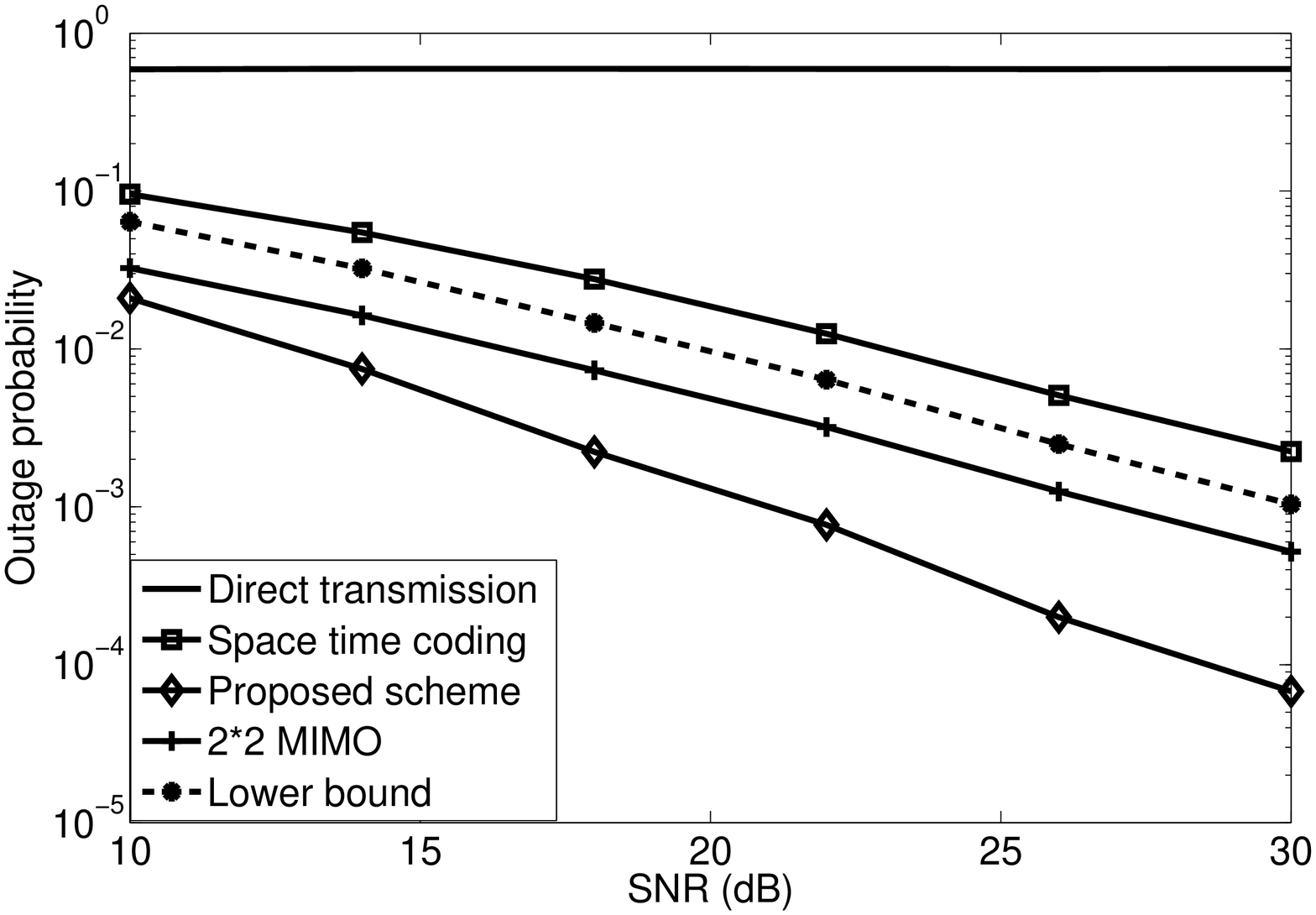}
\label{r1}} \hfil
\subfigure[$r=1.5$.]{\includegraphics[width=3.2in]{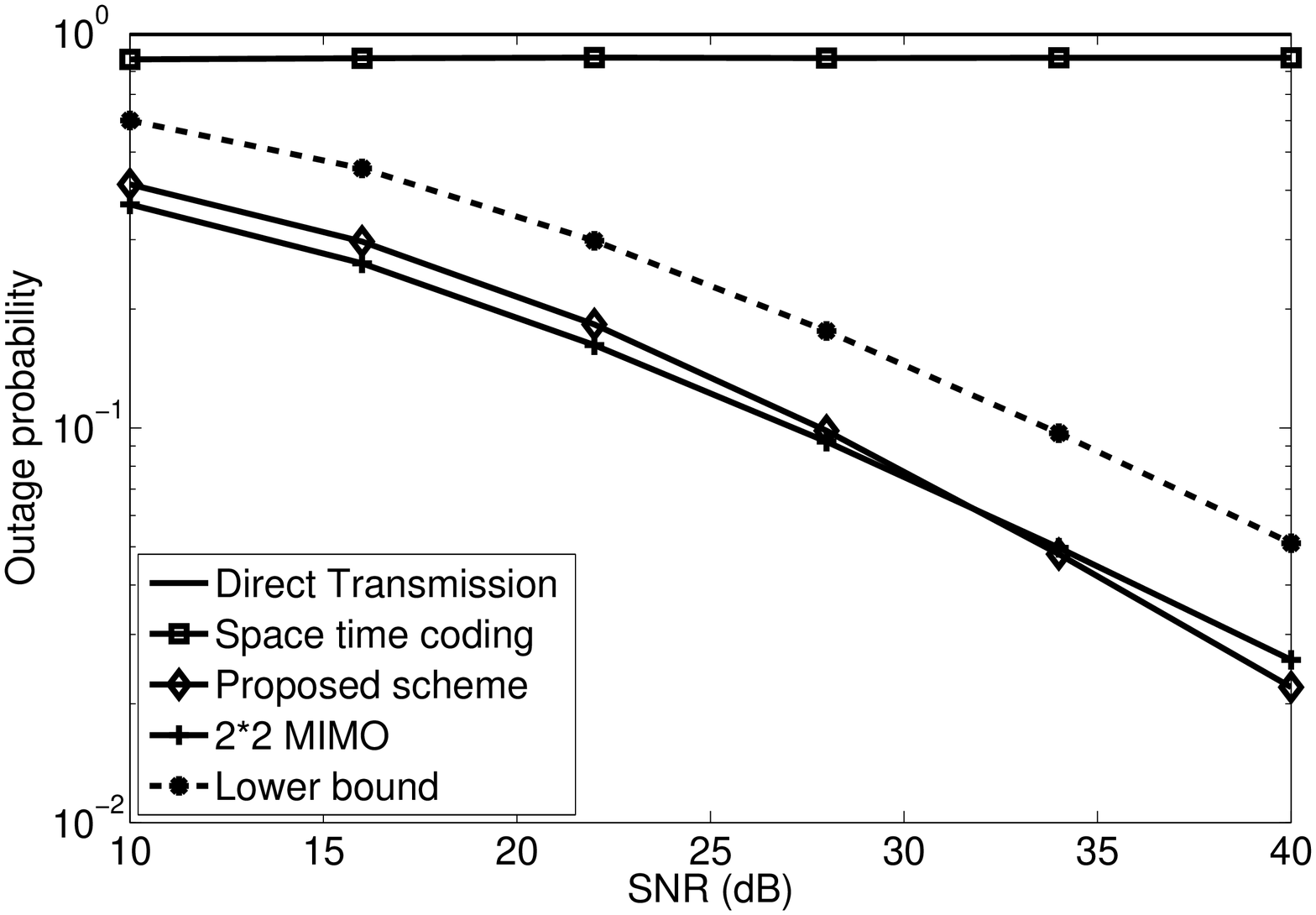}
\label{r15}} \caption{Outage probability for different schemes, when
(a) $r=1$, (b) $r=1.5$} \label{fsnr}
\end{figure}

It is not straightforward to analyze the finite-SNR DMT for
space-time coding. However, we expect from the conclusion in
\emph{Theorem 5} that space-time coding cannot perform better than
the proposed successive relaying scheme for $r>1$. Fig. \ref{fsnr},
which reflect the finite-SNR DMT properties for different schemes,
shows simulated values of the outage probability for different
values of the multiplexing gain $r$ when $L=20$, while assuming the
source-relay link is perfect. The ``lower bound'' curve represents
the performance lower bound (\ref{lob}) for the proposed successive
relaying scheme. Note that this is also a performance lower bound
for the distributed D-BLAST scheme. We can observe from the figures
that the lower bound is actually not very tight. The proposed scheme
in fact offers much better performance. This might be because the
successive relaying scheme can in fact offer a higher
\emph{diversity gain} than $2 \times 2$ MIMO transmission (see the
differences among the curves' slopes), as it uses three antennas
instead of two antennas to transmit. It can be seen that for $r=1$,
the diversity gain for direct transmission is zero and so the curve
has no slope. When $r=3/2$, both direct transmission and space-time
coding schemes have a diversity gain of zero, while the proposed
scheme has a diversity gain at least as good as that of a $2 \times
2$ MIMO system. This confirms the analysis in the paper and shows
that the proposed successive relaying scheme can offer significant
performance advantages over other schemes for higher data rates.

\section{Conclusions}

Unlike most previous work in wireless relay networks, which
concentrates only on cooperative diversity, this work opens a new
direction of exploiting the \emph{cooperative multiplexing} in such
networks. The results in the paper also suggest that distributed
coding cannot only offer diversity, but also multiplexing gain as
well. This discovery also implies a new direction for future network
coding design. A number of interesting topics are left for future
work: (a) exploiting more multiplexing benefits in the model
discussed in the paper; (b) extending the model to more general
cases in which every node is equipped with multiple antennas, or in
which there are multiple relays; (c) extending the analyses to a
multi-user environment, e.g, a multiple-access relay network; and
(d) exploiting the possibility and constraints of obtaining
multiplexing gain by using other relaying modes, such as
compress-and-forward.


\section*{Acknowledgment}
This research was supported in part by the U.S. National Science
Foundation under Grants ANI-03-38807 and CNS-06-25637. The authors
would like to thank Mr. Chao Wang for many helpful discussions
during the completion of this work.



%

\end{document}